\documentstyle[12pt]{article}
\def\fnote#1#2{\begingroup\def\thefootnote{#1}\footnote{#2}
\endgroup}

\begin{document}

\hfill UTTG-05-97

\begin{center}
{\bf What is Quantum Field Theory, and What Did We Think It
Is?}\fnote{*}{Talk presented at the conference ``Historical
and Philosophical Reflections on the Foundations of Quantum
Field Theory,'' at Boston University, March 1996.  It will
be
published in the proceedings of this conference.}

\vspace{12pt}
\noindent
Steven Weinberg\fnote{**}{Research supported in part by the
Robert A. Welch
 Foundation and NSF Grant PHY 9511632.  E-mail address:
weinberg@physics.utexas.edu}\\{}

\vspace{12pt}
\noindent
Physics Department, University of Texas at Austin\\
Austin, TX 78712
\end{center}

\vspace{18pt}

Quantum field theory was originally thought to be simply the
quantum theory of fields.  That is, when quantum mechanics
was developed physicists already knew about various
classical fields, notably the electromagnetic field, so what
else would they do but quantize the electromagnetic field in
the same way that they quantized the theory of single
particles?  In 1926, in one of the very first papers on
quantum mechanics,$^1$ Born, Heisenberg and Jordan presented
the
quantum theory of the electromagnetic field.  For simplicity
they left out the polarization of the photon, and took
spacetime to have one space and one time dimension, but that
didn't affect the main results.  ({\it Response to comment
from audience}:
Yes, they were really doing string theory, so in this sense
string theory is earlier than quantum field theory.)  Born
et
al. gave a formula for the electromagnetic field as a
Fourier transform and used the canonical commutation
relations to identify the coefficients in this Fourier
transform as operators that destroy and create photons, so
that when quantized this field theory became a theory of
photons.  Photons, of course, had been around (though not
under that name) since Einstein's work on the photoelectric
effect two decades earlier, but this paper showed that
photons are an inevitable consequence of quantum mechanics
as applied to electromagnetism.

The quantum theory of particles like electrons was being
developed at the same time, and made relativistic by
Dirac$^2$
in 1928--1930.  For quite a long time many physicists
thought that the world consisted of both fields and
particles:  the electron is a particle, described by a
relativistically invariant version of the Schr\"odinger wave
equation,  and the electromagnetic field is a field, even
though it also behaves like particles.  Dirac I think never
really changed his mind about this, and I believe that this
was Feynman's understanding when he first developed the path
integral and worked out his rules for calculating in quantum
electrodynamics.  When I first learned about the
path-integral formalism, it was in terms of
electron trajectories (as it is also presented in the book
by Feynman and Hibbs$^3$).  I already thought that wasn't
the
best way to look at electrons, so this gave me an
distaste for the path integral formalism, which
although unreasonable lasted until I learned
of 't Hooft's work$^4$ in 1971.  I feel it's all right to
mention autobiographical details like that as long as the
story shows how the speaker was wrong.

In fact, it was quite soon after the
Born--Heisenberg--Jordan paper of 1926 that the idea came
along that in fact
one could use quantum field theory for everything, not just
for electromagnetism.  This was the work of many theorists
during the period 1928--1934, including Jordan, Wigner,
Heisenberg, Pauli, Weisskopf, Furry, and Oppenheimer.
Although this is often talked about as second quantization,
I would like to urge that this description should be banned
from physics, because a quantum field is not a quantized
wave function.  Certainly the Maxwell field is not the wave
function of the photon, and for reasons that Dirac himself
pointed out, the Klein--Gordon fields that we use for pions
and Higgs bosons could not be the wave functions of the
bosons.  In its mature form, the idea of quantum field
theory is that quantum fields are the basic ingredients of
the universe, and particles are just bundles of energy and
momentum of the fields.   In a relativistic theory the  wave
function is a functional of these fields, not a function of
particle coordinates.  Quantum field theory hence led to a
more unified view of nature than the old dualistic
interpretation in terms of both fields and particles.

There is an irony in this.  (I'll point out several ironies
as I go along --- this whole subject is filled with
delicious ironies.)  It is that although the battle is over,
and the old dualism that treated photons in an entirely
different way from electrons is I think safely dead and will
never return, some calculations are actually easier in the
old particle framework.  When Euler, Heisenberg and
Kockel$^5$
in the mid-thirties calculated the effective action (often
called the Euler--Heisenberg action) of a constant external
electromagnetic field, they calculated to all orders in the
field, although their result is usually  presented only to
fourth order.  This calculation would probably have been
impossible with the old fashioned perturbation theory
techniques of the time, if they had not done it by first
solving the Dirac equation in a constant external
electromagnetic field and using those Dirac wave functions
to figure out the effective action.  These techniques of
using particle trajectories  rather than field histories in
calculations have been revived in recent years. Under the
stimulus of string theory, Bern and Kosower,$^6$ in
particular,
have developed a useful formalism for doing calculations by
following particle world lines rather than by thinking of
fields evolving in time.  Although this approach was
stimulated by string theory, it has been reformulated
entirely within the scope of ordinary quantum field theory,
and simply represents a more efficient way of doing certain
calculations.

One of the key elements in the triumph of quantum field
theory was the development of renormalization theory.  I'm
sure this has been discussed often here, and so I won't
dwell on it.  The version of renormalization theory that had
been developed in the late 1940s remained somewhat in the
shade for a long time for two reasons:  (1) for the weak
interactions it did not seem possible to develop a
renormalizable theory, and (2) for the strong interactions
it was easy to write down renormalizable theories, but since
perturbation theory was inapplicable it did not seem that
there was anything that could be done with these theories.
Finally all these problems were resolved through the
development of the standard model, which was triumphantly
verified by experiments during the mid-1970s, and today the
weak, electromagnetic and strong interactions are happily
all described by a renormalizable quantum field theory.  If
you had asked me in the mid-1970s about the shape of future
fundamental physical theories, I would have guessed that
they would take the form of better, more all-embracing, less
arbitrary, renormalizable quantum field theories.  I gave a
talk at the Harvard Science Center at around this time,
called ``The Renaissance of Quantum Field Theory,'' which
shows you the mood I was in.

There were two things that especially attracted me to the
ideas of renormalization and quantum field theory.  One of
them was that the requirement that a physical theory be
renormalizable is a precise and rational criterion of
simplicity.  In a sense, this requirement had been used long
before the advent of renormalization theory.  When Dirac
wrote down the Dirac equation in 1928 he could have added an
extra `Pauli' term$^7$ which would have given the electron
an
arbitrary anomalous magnetic moment.  Dirac could (and
perhaps did) say `I won't add this term because it's ugly
and complicated and there's no need for it.'   I think that
in physics this approach generally makes good strategies but
bad rationales.  It's often a good strategy to study simple
theories before you study complicated theories because it's
easier to see how they work, but the purpose of physics is
to find out why nature is the way it is, and simplicity by
itself is I think never the answer.  But renormalizability
was a condition of simplicity which was being imposed for
what seemed after Dyson's 1949 papers$^8$ like a rational
reason, and it explained not only why the electron has the
magnetic moment it has, but also (together with gauge
symmetries) all the detailed features of the standard model
of weak, electromagnetic, and strong, interactions, aside
from some numerical parameters.

The other thing I liked about quantum field theory during
this period of tremendous optimism was that it offered a
clear answer to the ancient question of what we mean by an
elementary particle: it is simply a particle whose field
appears in the Lagrangian. It doesn't matter if it's stable,
unstable, heavy, light --- if its field appears in the
Lagrangian then it's elementary, otherwise it's
composite.\fnote{***}{We should not really give quantum
field
theory too much credit for clarifying the distinction
between elementary and composite particles, because some
quantum field theories exhibit the phenomenon of
bosonization:  At least in two dimensions there are theories
of elementary scalars that are equivalent to theories with
elementary fermions.}

Now my point of view has changed.  It has changed partly
because of my  experience in teaching quantum field theory.
When you teach any branch of physics you must motivate the
formalism --- it isn't any good just to present the
formalism and say that it agrees with experiment --- you
have to explain to the students why this the way the world
is.  After all, this is our aim in physics, not just to
describe nature, but to explain nature.  In the course of
teaching quantum field theory, I developed a rationale for
it, which very briefly is that it is the only way of
satisfying the principles of Lorentz invariance plus quantum
mechanics plus one other principle.

Let me run through this argument very rapidly.  The first
point is to start with Wigner's definition of physical
multi-particle states as representations of the
inhomogeneous Lorentz group.$^9$  You then define
annihilation
and creation operators $a(\vec{p},\sigma,n)$ and
$a^\dagger(\vec{p},\sigma,n)$ that act on these states
(where $\vec{p}$ is the three-momentum, $\sigma$ is the spin
$z$-component, and $n$ is a species label).   There's no
physics in introducing such operators, for it is easy to see
that any operator whatever can be expressed as a functional
of them.  The existence of a Hamiltonian follows from
time-translation invariance, and much of physics is
described by
the $S$-matrix, which is given by the well known
Feynman--Dyson series of integrals over time of time-ordered
products of the interaction Hamiltonian $H_I(t)$  in the
interaction picture;
\begin{eqnarray} S&=&\sum_{n=0}^\infty\frac{(-
i)^n}{n!}\int_{-\infty}^\infty dt_1
\int_{-\infty}^\infty dt_2\cdots \int_{-\infty}^\infty dt_n
\nonumber\\&&\times\;
T\{H_I(t_1)H_I(t_2)\cdots H_I(t_n)\}\;.
\end{eqnarray}
This should all be familiar.    The other principle that has
to be added is the cluster decomposition principle, which
requires that distant experiments give uncorrelated
results.$^{10}$
In order to have cluster decomposition, the Hamiltonian is
written not just as any functional of creation and
annihilation operators, but as a power series in these
operators with coefficients that (aside from a {\em single}
momentum-conservation delta function) are sufficiently
smooth functions of the momenta carried by the operators.
This condition is satisfied for an interaction Hamiltonian
of the  form
\begin{equation}
H_I(t)=\int d^3x\; {\cal H}(\vec{x},t)
\end{equation}
where ${\cal H}(x)$ is a power series (usually a polynomial)
with terms that are local in annihilation fields, which are
Fourier transforms of the annihilation operators:
\begin{equation} \psi^{(+)}_\ell(x)= \int
d^3p\;\sum_{\sigma,n} e^{ip\cdot x }u_\ell({\vec
p},\sigma,n)\,a({\vec p},\sigma,n)
\end{equation}
together of course with their adjoints, the creation fields.

So far this all applies to nonrelativistic as well as
relativistic theories.\fnote{\dagger}{By the way, the reason
that
quantum field theory is useful even in nonrelativistic
statistical mechanics, where there is often a selection rule
that makes the actual creation or annihilation of particles
impossible, is that in statistical mechanics you have to
impose a cluster decomposition principle, and quantum field
theory is the natural way to do so.}  Now if you also want
Lorentz invariance, then you have to face the fact that the
time-ordering in the Feynman--Dyson series (1) for the
$S$-matrix doesn't look very Lorentz invariant.  The obvious
way
to make the $S$-matrix Lorentz invariant is to take the
interaction Hamiltonian density ${\cal H}(x)$ to be a
scalar, and also to require that these Hamiltonian densities
commute at spacelike separations
\begin{equation}
[{\cal H}(x),{\cal H}(y)]=0~~{\rm for~spacelike}~~x-y\;,
\end{equation}
in order to exploit the fact that time ordering {\em is}
Lorentz invariant when the separation between spacetime
points is timelike.  In order to satisfy the requirement
that the Hamiltonian density commute with itself at
spacelike separations, it is constructed out of fields which
satisfy the same requirement.  These are given by sums of
fields that annihilate particles plus fields that create the
corresponding antiparticles
\begin{eqnarray}
\psi_\ell(x)&=&\sum_{\sigma,n}\int d^3p\;\Bigg[
e^{ip\cdot x}\,u_\ell({\vec p},\sigma,n)\,a({\vec
p},\sigma,n)\nonumber\\&&+e^{-ip\cdot x}\,v_\ell({\vec
p},\sigma,n)\,a^\dagger({\vec p},\sigma,\bar{n})\Bigg]\;,
\end{eqnarray}
where $\bar{n}$ denotes the antiparticle of the particle of
species $n$.
For a field $\psi_\ell$ that transforms according to an
irreducible representation of the homogeneous Lorentz group,
the form of the coefficients $u_\ell$ and $v_\ell$
is completely determined (up to a single over-all constant
factor) by the Lorentz transformation properties of the
fields and one-particle states, and by the condition that
the fields commute at spacelike separations.
Thus the whole formalism of fields, particles, and
antiparticles seems to be an inevitable consequence of
Lorentz invariance, quantum mechanics, and cluster
decomposition, without any ancillary assumptions about
locality or causality.

This discussion has been extremely sketchy, and is subject
to all sorts of qualifications.  One of them is that for
massless particles, the range of possible theories is
slightly larger than I have indicated here.  For example, in
quantum electrodynamics, in a physical gauge like Coulomb
gauge, the Hamiltonian is not of the form (2) --- there is
an additional term, the Coulomb potential, which is  bilocal
and serves to cancel a non-covariant term in the propagator.
But relativistically invariant quantum theories always (with
some qualifications I'll come to later) do turn out to be
quantum field theories, more or less as I have described
them here.

One can go further, and ask why we should formulate our
quantum field theories in terms of Lagrangians.    Well, of
course creation and annihilation operators by themselves
yield pairs of canonically conjugate  variables; from the
$a$s and $a^\dagger$s, it is easy to  construct $q$s and
$p$s.  The time-dependence of these operators is dictated in
terms of the Hamiltonian, the generator of time
translations, so the Hamiltonian formalism is trivially
always with us.  But why the Lagrangian formalism?  Why do
we enumerate possible theories by giving their Lagrangians
rather than by writing down Hamiltonians?  I think the
reason for this is that it is only in the Lagrangian
formalism (or more generally the action formalism)  that
symmetries imply the existence of Lie algebras of suitable
quantum operators, and you need these Lie algebras to make
sensible quantum theories.  In particular, the $S$-matrix
will be Lorentz invariant if there is a set of 10
sufficiently smooth operators satisfying the commutation
relations of the inhomogeneous Lorentz group.  It's not
trivial to write down a Hamiltonian that will give you a
Lorentz invariant $S$-matrix --- it's not so easy to think
of the Coulomb potential just on the basis of Lorentz
invariance --- but if you start with a Lorentz invariant
Lagrangian density then because of Noether's theorem the
Lorentz invariance of the $S$-matrix is automatic.

Finally, what is the motivation for  the special gauge
invariant Lagrangians that we use in the standard model and
general relativity?  One possible answer is that quantum
theories of  mass zero, spin one particles violate Lorentz
invariance unless the fields are coupled in a gauge
invariant way, while quantum theories of mass zero, spin two
particles violate Lorentz invariance unless the fields are
coupled in a way that satisfies the equivalence principle.

This has been an outline of the way I've been teaching
quantum field theory these many years.  Recently I've put
this all together into a book,$^{11}$ now being sold for a
negligible price.   The bottom line is that
quantum mechanics plus Lorentz invariance plus cluster
decomposition implies quantum field theory.  But there are
caveats that have to be attached to this, and I can see
David Gross in the front row anxious to take me by the
throat over various gaps in what I have said, so I had
better list these caveats quickly to save myself.

First of all, the argument I have presented is obviously
based on perturbation theory.  Second, even in perturbation
theory, I haven't stated a clear theorem, much less proved
one.  As I mentioned there are complications when you have
things like mass zero, spin one particles for example; in
this case you don't really have a fully Lorentz invariant
Hamiltonian density, or even one that is completely local.
Because of these complications, I don't know how even to
state a general theorem, let alone prove it, even in
perturbation theory.  But I don't think that these are
insuperable obstacles.

A much more serious objection to this
not-yet-formulated theorem is that there's already a counter
example to it: string theory.  When you first learn string
theory it seems in an almost miraculous way to give Lorentz
invariant, unitary $S$-matrix elements without being a field
theory in the sense that I've been using it.  (Of course it
is a field theory in a different sense --- it's a two
dimensional conformally invariant field theory, but not a
quantum field theory in four spacetime dimensions.)  So
before even being formulated precisely, this theorem suffers
from at least one counter example.

Another fundamental problem is that the $S$-matrix isn't
everything.  Spacetime could be radically curved, not just
have little ripples on it.  Also, at finite temperature
there's no $S$-matrix because particles cannot get out to
infinite distances from a collision without bumping into
things.  Also, it seems quite possible that at very short
distances the description of events in four-dimensional flat
spacetime becomes inappropriate.

Now, all of these caveats really work only against the idea
that the final theory of nature is a quantum field theory.
They leave open the view, which is in fact the point of view
of my book,  that although you can not argue that relativity
plus quantum mechanics plus cluster decomposition
necessarily leads only to quantum field theory,   it is very
likely that any quantum theory that at sufficiently low
energy and large distances looks Lorentz invariant and
satisfies the cluster decomposition principle  will also at
sufficiently low energy {\em look} like a quantum field
theory.      Picking up a phrase from Arthur Wightman, I'll
call this a folk theorem.  At any rate, this folk theorem is
satisfied by string theory, and we don't know of any
counterexamples.

This leads us to the idea of effective field theories.  When
you use  quantum field theory to study low-energy phenomena,
then according to the folk theorem you're not really making
any assumption that could be wrong, unless of course Lorentz
invariance or quantum mechanics or cluster decomposition is
wrong, provided you don't say specifically what the
Lagrangian is. As long as you let it be the most general
possible Lagrangian consistent with the symmetries of the
theory, you're simply writing down the most general theory
you could possibly write down.  This point of view has been
used in the last fifteen years or so to justify the use of
effective field theories, not just  in the tree
approximation where they had been used for some time
earlier, but also including loop diagrams.  Effective field
theory was first used in this way to calculate processes
involving soft $\pi$ mesons,$^{12}$ that is, $\pi$ mesons
with energy less
than about $2\pi F_\pi\approx 1200$ MeV.  The use of
effective quantum field theories has been extended more
recently to nuclear physics,$^{13}$  where although nucleons
are
not soft they never get far from their mass shell, and for
that reason can be also treated by similar methods as the
soft pions.  Nuclear physicists have adopted this point of
view, and I gather that they are happy about using this new
language because it allows one to show in a fairly
convincing way that what they've been doing all along (using
two-body potentials only, including one-pion exchange and a
hard core) is the correct first step in a consistent
approximation scheme.      The effective field theory
approach has been applied more recently to
superconductivity. Shankar, I believe, in a contribution to
this conference is talking about this.  The present educated
view of the standard model, and of general
relativity,$^{14}$ is
again that these are the leading terms in effective field
theories.

The essential point in using  an effective field theory is
you're not allowed to make any assumption of simplicity
about the Lagrangian.  Certainly you're not allowed to
assume renormalizability.  Such assumptions might be
appropriate if you were dealing with a fundamental theory,
but not for an effective field theory, where you must
include all possible terms that are consistent with the
symmetry.  The thing that makes this procedure useful is
that although the more complicated terms are not excluded
because they're non-renormalizable, their effect is
suppressed by factors of the ratio of the energy to some
fundamental energy scale of the theory.   Of course,  as you
go to higher and higher energies, you have more and more of
these suppressed terms that you have to worry about.

On this basis, I don't see any reason why anyone today would
take Einstein's general theory of relativity seriously as
the foundation of a quantum theory of gravitation, if by
Einstein's theory is meant the theory with a Lagrangian
density given by just the term $\sqrt{g}R/16\pi G$.  It
seems to me there's no reason in the world to suppose that
the Lagrangian does not contain all the higher terms with
more factors of the curvature and/or more derivatives, all
of which are suppressed by inverse powers of the Planck
mass, and of course don't show up at any energy far below
the Planck mass, much less in astronomy or particle physics.
Why would anyone suppose that these higher terms are absent?

Likewise,  since now we know that without new fields there's
no way that the renormalizable terms in the standard model
could violate baryon conservation or lepton conservation, we
now understand in a rational way why baryon number and
lepton number are as well conserved as they are, without
having to assume that they are exactly
conserved.\fnote{\dagger\dagger}{The extra fields required
by low-energy
supersymmetry may invalidate this argument.}  Unless someone
has some {\em a priori} reason for exact baryon and lepton
conservation of which I haven't heard, I would bet very
strong odds that baryon number and lepton number
conservation are in fact violated by suppressed
non-renormalizable corrections to the standard model.

These effective field theories are
non-renormalizable in the old Dyson power-counting sense.
That is, although to achieve a given accuracy at any given
energy,  you need only take account of a finite number of
terms in the action, as you increase the accuracy or the
energy you need to include more and more terms,  and so you
have to know more and  more.     On the other hand,
effective field theories still must be renormalizable
theories in what I call the modern sense:  the symmetries
that govern the action also have to govern the infinities,
for otherwise there will be infinities that can't be
eliminated by absorbing them into counter terms to the
parameters in the action.      This requirement is
automatically satisfied for unbroken global symmetries, such
as Lorentz invariance and isotopic spin invariance and so
on. Where it's not trivial is for gauge symmetries. We
generally deal with gauge theories by choosing a gauge
before quantizing the theory, which of course breaks the
gauge invariance, so it's not obvious how gauge invariance
constrains the infinities.   (There is a symmetry called
BRST invariance$^{15}$ that survives gauge fixing, but
that's non-linearly realized, and non-linearly realized
symmetries of
the action are not symmetries of the Feynman amplitudes.)
This raises a question, whether gauge theories that are not
renormalizable in the power counting sense are
renormalizable in the modern sense.   The theorem that says
that infinities are governed by the same gauge symmetries as
the terms in the Lagrangian was originally proved back in
the old days by 't Hooft and Veltman$^{16}$ and Lee and
Zinn-Justin$^{17}$
only for theories that are renormalizable in the old
power-counting sense, but this theorem has only recently
been
extended to theories of the Yang--Mills$^{18}$ or Einstein
type
with arbitrary numbers of complicated interactions that are
not renormalizable in the power-counting
sense.\fnote{\ddagger}{I
refer here to work of myself and Joaquim Gomis,$^{19}$
relying on
recent theorems about the cohomology of the
Batalin--Vilkovisky operator by Barnich, Brandt, and
Henneaux.$^{20}$
Earlier work along these lines but with different motivation
was done by Voronov, Tyutin, and Lavrov;$^{21}$
Anselmi;$^{22}$ and
Harada, Kugo, and
Yamawaki.$^{23}$  }   You'll be reassured to know that these
theories are renormalizable in the modern sense, but there's
no proof that this will be true of all quantum field
theories with local symmetries.

I promised you a few ironies today. The second one takes me
back to the early 1960s when $S$-matrix theory was very
popular at Berkeley and elsewhere.  The hope of $S$-matrix
theory was that, by using the principles of unitarity,
analyticity, Lorentz invariance and other symmetries, it
would be possible to calculate the $S$-matrix, and you would
never have to think about a quantum field. In a way, this
hope reflected a kind of positivistic puritanism: we can't
measure the field of a pion or a nucleon, so we shouldn't
talk about it, while we do measure $S$-matrix elements, so
this is what we should stick to as ingredients of our
theories. But more important than any philosophical hang-ups
was the fact that  quantum field theory  didn't seem to be
going anywhere in accounting for the strong and weak
interactions.

One problem with the $S$-matrix program was in formulating
what is meant by the analyticity of the $S$-matrix.    What
precisely are the analytic properties of a multi-particle
$S$-matrix element?  I don't think anyone ever knew.  I
certainly didn't know, so even though I was at Berkeley I
never got too enthusiastic about the details of this
program, although I thought it was a lovely idea in
principle. Eventually the $S$-matrix program had to retreat,
as described by Kaiser in a contribution to this conference,
to a sort of mix of field theory and $S$-matrix theory.
Feynman rules were used to find the singularities in the
$S$-matrix, and then they were thrown away, and the analytic
structure of the $S$-matrix with these singularities,
together with unitarity and Lorentz invariance, was used to
do calculations.

Unfortunately to use these assumptions it was necessary to
make uncontrolled approximations, such as the strip
approximation, whose mention will bring tears to the eyes of
those of us who are old enough to remember it.  By the
mid-1960's it was clear that $S$-matrix theory had failed in
dealing with the one problem it had tried hardest to solve,
that of pion--pion scattering.  The strip approximation
rested on the assumption that double dispersion relations
are dominated by regions of the Mandelstam diagram near the
fringes of the physical region, which  would only make sense
if $\pi$--$\pi$ scattering is strong at low energy, and
these calculations  predicted that $\pi$--$\pi$ scattering
is indeed strong at low energy, which was at least
consistent, but it was then discovered that $\pi$--$\pi$
scattering is {\em not} strong at low energy.  Current
algebra came along at just that time, and was used to
predict not only that low energy $\pi$-$\pi$ scattering is
not strong, but also successfully predicted the values of
the $\pi$--$\pi$ scattering lengths.$^{24}$ From a practical
point
of view, this was the greatest defeat of $S$-matrix theory.
The irony here is that the $S$-matrix philosophy is not that
far from the modern philosophy of effective field theories,
that what you should do is just write down the most general
$S$-matrix  that satisfies basic principles.  But the
practical way to implement $S$-matrix theory is to use an
effective quantum field theory --- instead of deriving
analyticity properties from Feynman diagrams, we use the
Feynman diagrams themselves.  So here's another answer to
the question of what quantum field theory is: it is
$S$-matrix theory, made practical.

By the way, I think that the emphasis in $S$-matrix theory
on analyticity as a fundamental principle was misguided, not
only because no one could ever state  the detailed
analyticity properties of general $S$-matrix elements, but
also because Lorentz invariance requires causality (because
as I argued earlier otherwise you're not going to get a
Lorentz invariant $S$-matrix), and in quantum field theory
causality allows you to derive analyticity properties.  So I
would include Lorentz invariance, quantum mechanics and
cluster decomposition as fundamental principles, but not
analyticity.

 As I have said, quantum field theories provide an expansion
in powers of the energy of a process divided by  some
characteristic energy; for soft pions this characteristic
energy is about a GeV; for superconductivity it's the Debye
frequency or temperature; for the standard model it's
$10^{15}$ to $10^{16}$ GeV; and for gravitation it's about
$10^{18}$ GeV.  Any effective field theory loses its
predictive power when the energy of the processes in
question approaches the characteristic energy.  So what
happen to the effective field  theories of electroweak,
strong, and gravitational interactions at  energies of order
$10^{15}$--$10^{18}$ GeV?  I know of only two plausible
alternatives.

One possibility is that the theory remains a quantum field
theory, but one in which  the finite or infinite number of
renormalized couplings do not run off to infinity with
increasing energy, but hit a fixed point of the
renormalization group equations.  One way that can happen
is provided by asymptotic freedom in a renormalizable
theory,$^{25}$ where the fixed point is at zero coupling,
but it's
possible to have more general fixed points with infinite
numbers of non-zero nonrenormalizable couplings.  Now, we
don't know how to calculate these non-zero fixed points very
well, but one thing we know with fair certainty is that the
trajectories that run into a fixed point in the ultraviolet
limit form a finite dimensional subspace of the infinite
dimensional space of all coupling constants.  (If anyone
wants to know how we know that, I'll explain this later.)
That means that the condition, that the trajectories hit a
fixed point, is just as restrictive in a nice way as
renormalizability used to be:  It  reduces the number of
free coupling parameters to a finite number.  We don't yet
know how to do calculations for fixed points that are not
near zero coupling.  Some time ago I proposed$^{26}$ that
these
calculations could be done in the theory of gravitation by
working in $2 + \epsilon$ dimensions and expanding in powers
of $\epsilon=2$, in analogy with the way  that Wilson and
Fisher$^{27}$ had calculated critical exponents by working
in
$4 - \epsilon$ dimensions and expanding in powers of
$\epsilon=1$, but this program doesn't seem to be working
very well.

 The other possibility, which I have to admit is {\em a
priori} more likely, is that at very high energy we will run
into  really new physics, not describable in terms of a
quantum field theory.  I think that by far the most likely
possibility is that this will be something like a string
theory.

Before I leave the renormalization group, I did want to say
another word about it because there's going to be an
interesting discussion on this subject here tomorrow
morning, and for reasons I've already explained I can't be
here.  I've read  a lot of argument  about the Wilson
approach$^{28}$ vs. the
Gell-Mann--Low approach,$^{29}$ which seems to me to call
for
reconciliation.  There have been two fundamental insights in
the development of the renormalization group.  One, due to
Gell-Mann and Low, is that logarithms of energy that violate
naive scaling and invalidate perturbation theory arise
because of the way that renormalized coupling constants are
defined, and that these logarithms can be avoided by
renormalizing at a sliding energy scale.  The  second
fundamental insight, due to Wilson, is that it's very
important in dealing with phenomena at a certain energy
scale to integrate out the physics at much higher energy
scales.  It seems to me these are the same insight, because
when you adopt the Gell-Mann--Low approach and define a
renormalized coupling at a sliding scale and use
renormalization theory to eliminate the infinities rather
than an explicit cutoff, you are in effect integrating out
the higher energy degrees of freedom --- the integrals
converge because after renormalization the integrand begins
to fall off rapidly at the energy scale at which the
coupling constant is defined.  (This is true whether or not
the theory is renormalizable in the power-counting sense.)
So in other words instead of a sharp cutoff {\em a la}
Wilson, you have a soft cutoff, but it's a cutoff
nonetheless and it serves the same purpose of integrating
out the short distance degrees of freedom.    There are
practical differences between the Gell-Mann--Low and Wilson
approaches, and there are some problems for which one is
better and other problems for which the other is better.  In
statistical mechanics it isn't important to maintain Lorentz
invariance, so you might as well have a cutoff.  In  quantum
field theories, Lorentz invariance is necessary, so it's
nice to renormalize {\em a la} Gell-Mann--Low. On the other
hand, in supersymmetry theories there are some
non-renormalization theorems that are simpler if you use a
Wilsonian cutoff than a Gell-Mann--Low cutoff.$^{30}$  These
are
all practical differences, which we have to take into
account, but I don't find any fundamental philosophical
difference between these two approaches.

On the plane coming here I read a comment by Michael
Redhead, in a paper submitted to this conference: `To
subscribe to the new effective field theory programme is to
give up on this endeavor' [the endeavor of finding really
fundamental laws of nature], `and retreat to a position that
is somehow less intellectually exciting.'   It seems to me
that this is analogous to saying that to balance your
checkbook is to give up dreams of wealth and have a life
that is intrinsically less exciting.  In a sense that's
true,  but nevertheless it's still something that
you had better do every once in a while.  I think that in
regarding the standard model and general relativity as
effective field theories we're simply balancing our
checkbook and realizing that we perhaps didn't know as much
as we thought we did,  but this is the way the world is and
now we're going to go on the next step and try to find an
ultraviolet fixed point, or (much more likely) find entirely
new physics.     I have said that I thought that this new
physics takes the form of  string theory,  but of course, we
don't know if that's the final answer.   Nielsen and
Oleson$^{31}$
showed long ago that relativistic quantum field theories can
have string-like solutions.   It's conceivable, although I
admit not entirely likely,  that something like modern
string theory arises from a quantum field theory.  And that
would be the final irony.

\pagebreak

\begin{center}
{\bf References}
\end{center}

\begin{enumerate}

\item M. Born, W. Heisenberg, and P. Jordan, {\it Z. f.
Phys.} {\bf 35}, 557 (1926).

\item P.A.M. Dirac, {\it Proc. Roy. Soc.}   {\bf A117}, 610
(1928); {\it ibid.}, {\bf A118}, 351 (1928); {\it ibid.},
{\bf A126}, 360 (1930).

\item R. P. Feynman and A. R. Hibbs, {\it Quantum Mechanics
and Path Integrals} (McGraw-Hill, New York, 1965).

\item G. 't Hooft, {\em Nucl. Phys.} {\bf B35}, 167 (1971).

\item H. Euler and B. Kockel, {\it Naturwiss.} {\bf 23}, 246
(1935); W. Heisenberg and H. Euler,  {\it Z. f. Phys.} {\bf
98}, 714 (1936).

\item Z. Bern and D. A. Kosower, in {\em International
Symposium on Particles, Strings, and Cosmology}, eds. P.
Nath and S. Reucroft (World Scientific, Singapore, 1992):
794; {\it Phys. Rev. Lett.} {\bf 66}, 669 (1991).

\item W. Pauli, {\it Z. f. Phys.} {\bf 37}, 263 (1926); {\bf
43}, 601 (1927).

\item F.J. Dyson, {\it Phys. Rev.} {\bf 75}, 486, 1736
(1949).

\item E. P. Wigner, {\it Ann. Math.} {\bf 40}, 149 (1939).

\item The cluster decomposition principle seems to have been
first stated explicitly in quantum field theory by E. H.
Wichmann and J. H. Crichton, {\it Phys. Rev.} {\bf 132},
2788 (1963).

\item S. Weinberg, {\em The Quantum Theory  of Fields ---
Volume I:
Foundations}~
(Cambridge University Press, Cambridge, 1995)

\item S. Weinberg, {\em Phys. Rev. Lett.} {\bf 18}, 188
(1967); {\it Phys. Rev.} {\bf 166}, 1568 (1968); {\it
Physica} {\bf 96A}, 327 (1979).

\item S. Weinberg, {\em Phys. Lett.} {\bf B251}, 288 (1990);
{\em Nucl. Phys.} {\bf B363}, 3 (1991);  {\em Phys. Lett.}
{\bf B295}, 114 (1992).  C. Ord\'o\~nez and U. van Kolck,
{\em Phys. Lett.} {\bf B291}, 459 (1992); C. Ord\'o\~nez, L.
Ray, and U. van Kolck, {\em Phys. Rev. Lett.} {\bf 72}, 1982
(1994); U. van Kolck, {\em Phys. Rev.}, {\bf C49}, 2932
(1994); U. van Kolck, J. Friar, and T. Goldman, to appear in
{\em Phys. Lett. B}.   This approach to nuclear forces is
summarized in C. Ord\'o\~nez, L. Ray, and U. van Kolck,
Texas preprint UTTG-15-95, nucl-th/9511380, submitted to
{\em Phys. Rev. C}; J. Friar,
{\em Few-Body Systems Suppl.} {\bf 99}, 1 (1996).   For
application of these techniques to related nuclear
processes, see T.-S. Park, D.-P. Min, and M. Rho, {\em Phys.
Rep.} {\bf 233}, 341 (1993); Seoul preprint SNUTP 95-043,
nucl-th/9505017; S.R. Beane, C.Y. Lee, and U. van Kolck,
{\em Phys. Rev.}, {\bf C52}, 2915 (1995); T. Cohen, J.
Friar, G. Miller, and U. van Kolck, Washington preprint
DOE/ER/40427-26-N95, nucl-th/9512036.

\item J. F. Donoghue, {\it Phys. Rev.} {\bf D 50}, 3874
(1994).

\item C. Becchi, A. Rouet, and R. Stora, {\it Comm. Math.
Phys.} {\bf 42}, 127 (1975); in {\em Renormalization
Theory}, eds. G. Velo and A. S. Wightman (Reidel, Dordrecht,
1976); {\it Ann. Phys.} {\bf 98}, 287 (1976); I. V. Tyutin,
Lebedev Institute preprint N39 (1975).

\item G. 't Hooft and M. Veltman, {\it Nucl.
Phys.} {\bf B50}, 318 (1972).

\item B. W. Lee and J. Zinn-Justin, {\it Phys.
Rev.} {\bf D5}, 3121, 3137 (1972); {\it Phys. Rev.} {\bf
D7}, 1049 (1972).

\item C. N. Yang and R. L. Mills, {\it Phys.~Rev.}~{\bf 96},
191 (1954).

\item J. Gomis and S. Weinberg, Nuclear Physics B {\bf 469},
475--487 (1996).

\item G. Barnich and M. Henneaux,
{\em Phys. Rev. Lett.} {\bf 72}, 1588 (1994); G. Barnich, F.
Brandt,  and M. Henneaux, {\em Phys. Rev.} {\bf 51}, R143
(1995); {\em Commun. Math. Phys.} {\bf 174}, 57, 93 (1995);
{\em Nucl. Phys.} {\bf B455}, 357 (1995).

\item B. L. Voronov and I. V. Tyutin, {\it Theor. Math.
Phys.} {\bf 50}, 218 (1982); {\bf 52}, 628 (1982); B. L.
Voronov, P. M. Lavrov, and I. V. Tyutin, {\em Sov. J. Nucl.
Phys.} {\bf 36}, 292 (1982); P. M. Lavrov and I. V. Tyutin
{\it Sov. J. Nucl. Phys.} {\bf 41}, 1049 (1985).

\item D. Anselmi, {\em Class. and Quant. Grav.} {\bf 11},
2181 (1994); {\bf 12}, 319 (1995).

\item M. Harada, T. Kugo, and K. Yamawaki, {\em Prog. Theor.
Phys.} {\bf 91}, 801 (1994).

\item S. Weinberg, {\em Phys. Rev. Lett.} {\bf 16}, 879
(1966).

\item D. J. Gross and F. Wilczek, {\it Phys. Rev. Lett.},
{\bf 30}, 1343 (1973); H. D. Politzer, {\it Phys. Rev.
Lett.}, {\bf 30}, 1346 (1973).

\item S. Weinberg, in  {\it General Relativity},  eds. S. W.
Hawking and W. Israel, eds. (Cambridge University Press,
Cambridge, 1979): p. 790.

\item K. G. Wilson and M. E.  Fisher, {\it Phys. Rev. Lett.}
{\bf 28}, 240 (1972); K. G. Wilson, {\it Phys. Rev. Lett.}
{\bf 28}, 548 (1972).

\item K. G. Wilson, {\it Phys. Rev.} {\bf B4}, 3174, 3184
(1971); {\it Rev. Mod. Phys.} {\bf 47}, 773 (1975).

\item M. Gell-Mann and F. E.  Low, {\em Phys. Rev.} {\bf
95}, 1300 (1954).

\item V. Novikov, M. A. Shifman, A. I. Vainshtein, and V. I.
Zakharov, {\it Nucl. Phys.}  {\bf B229}, 381 (1983); M. A.
Shifman and  A. I. Vainshtein,  {\it Nucl. Phys.}  {\bf
B277}, 456 (1986); and references quoted therein.  See also
M. A. Shifman and  A. I. Vainshtein,  {\it Nucl. Phys.}
{\bf B359}, 571 (1991).

\item H. Nielsen and P. Oleson, {\em Nucl. Phys.} {\bf B61},
45 (1973).

\end{enumerate}
\end{document}